\documentstyle[aps,pra,preprint]{revtex}
\tighten

\begin{document}
\title{Optimal quantum teleportation with an arbitrary pure state}
\author{Konrad Banaszek}
\address{Rochester~Center~for~Quantum~Information and
Rochester~Theory~Center for Optical Science and Engineering,
University~of~Rochester, Rochester NY 14627\\
and Instytut Fizyki Teoretycznej, Uniwersytet Warszawski, Ho\.{z}a 69,
PL--00--681 Warszawa, Poland}
\date{\today}

\maketitle
\draft

\begin{abstract}
We derive the maximum fidelity
attainable for teleportation using a shared pair of $d$-level
systems in an arbitrary pure state. This derivation provides a complete
set of necessary and sufficient conditions for optimal 
teleportation protocols. We also discuss the information
on the teleported particle which is revealed in course of the
protocol using a non-maximally entangled state.
\end{abstract}

Entanglement is a key ingredient of quantum techniques for 
information processing.  One of the striking consequences
of quantum entanglement is the existence of the procedure called quantum
teleportation \cite{BennBrasPRL93}. This procedure allows two distant
parties, traditionally called Alice and Bob, to transmit faithfully
the quantum state of a particle. The resources needed for this purpose is
a pair of particles in a maximally entangled state shared by Alice and
Bob, and the possibility to transmit classical messages from Alice to Bob.
The teleportation procedure is an extremely useful tool for understanding
many properties of quantum entanglement \cite{PlenVedrCoP98}.

An important problem in quantum information theory is the
characterization of the entanglement exhibited by general quantum
states of bipartite systems, and the evaluation of their capability to
perform various quantum information processing
tasks.  In this paper we consider
the following question.  Suppose that Alice wants to teleport to Bob an
unknown pure state $|\psi\rangle_{1}$ of a $d$-level particle. Alice
and Bob share a single pair of $d$-level particles in a pure state
$|\text{tele}\rangle_{23}$. What is the maximum fidelity of teleportation
using such a state, and what teleportation protocols achieve this limit?

We present a derivation of the maximum fidelity, which defines a set of
necessary and sufficient conditions for a given protocol to be
optimal.  These conditions turn out to be satisfied by the standard
teleportation protocol based on projections on maximally entangled states
with appropriately adjusted Schmidt bases. 
The problem of optimal teleportation can be related to local transformations
of entangled states \cite{Monotones}:
as shown by the Horodecki's, there is a simple algebraic link between the
optimal teleportation fidelity and the maximum singlet
fraction \cite{HoroHoroPRA99}. Thus, our result provides also the maximum
singlet fraction for an arbitrary pure state of two $d$-level systems.

Of course, use of a non-maximally entangled state makes the
teleportation procedure imperfect. Nevertheless, we demonstrate
in this paper that one
can find a silver lining in such a case: namely, that the teleportation
procedure reveals some information on the teleported quantum state.
This information
can be converted into an estimate of the quantum state of the particle
initially possessed by Alice. We derive here an upper bound for the mean
estimation fidelity \cite{OptimalEstimation}, and provide an explicit
recipe for constructing the quantum state estimate that
saturates this bound.

We shall consider a general teleportation strategy, consisting of
an arbitrary measurement performed on Alice's side, followed by a general
transformation of Bob's particle. In the most general case,
Alice's measurement is
described by a certain positive operator-valued measure. Such a measure
can be decomposed into rank one operators, which are represented
by projections on not necessarily normalized states $|\Phi_r \rangle_{12}
\langle \Phi_r|$, where the index $r$ runs over all possible outcomes
of Alice's measurement. The unnormalized state vector of the particle
owned by Bob, after Alice has measured the outcome $r$, is given by:
\begin{equation}
|b_r \rangle_3 = {}_{12}\langle \Phi_r |(|\psi\rangle_{1}
\otimes |\text{tele}\rangle_{23}).
\end{equation}
After having received from Alice the outcome of her measurement,
Bob performs a general transformation of his particle, described
by:
\begin{equation}
\label{Eq:Brs}
|b_r \rangle_{3} \langle b_r |
\rightarrow
\sum_{s} \hat{B}_{rs} |b_r \rangle_{3} \langle b_r | \hat{B}^\dagger_{rs}
,
\end{equation}
where the operators $\hat{B}_{rs}$ satisfy
\begin{equation}
\sum_{s} \hat{B}^\dagger_{rs} \hat{B}_{rs} = \hat{\openone}
\end{equation}
for each $r$. In order to simplify the notation, we shall not write
explicitly the range of the parameter $s$, which can be different for
various values of $r$.

We shall quantify the quality of teleportation with
the help of the mean fidelity.
The probability that Alice obtains from her measurement the outcome $r$
is given by the scalar product ${}_3\langle b_r | b_r \rangle_3$.
The normalized state held by the Bob in this case is
$|b_r\rangle_3 / \sqrt{{}_3\langle b_r | b_r \rangle_3}$. After the
transformation of this state described by Eq.~(\ref{Eq:Brs}), 
its overlap with the original state vector $|\psi\rangle$ is given
by $\sum_s | {}_3 \langle \psi | \hat{B}_{rs} |b_r \rangle_3 |^2
/ {}_3\langle b_r | b_r \rangle_3$. Summation of this expression
over $r$ with the weights ${}_3\langle b_r | b_r \rangle_3$,
and integration over all possible input states $|\psi\rangle$,
yields the complete expression for the mean fidelity:
\begin{equation}
\bar{f} = \int\text{d} \psi \sum_{rs} \left| ({}_{12}\langle \Phi_r
| \otimes {}_{3}\langle
\psi |) \hat{B}_{rs} (|\psi\rangle_{1} \otimes | \text{tele} \rangle_{23})
\right|^2,
\end{equation}
where the integral $\int \text{d}\psi$ over the space of pure states
is performed using the canonical measure invariant with respect to
unitary transformations of the states $|\psi\rangle$.

Let us now select in the Hilbert spaces of the particles 2 and 3 the
orthonormal bases defined by the Schmidt decomposition of the 
shared state $|\text{tele}\rangle_{23}$:
\begin{equation}
|\text{tele}\rangle_{23} = \sum_{k=0}^{d-1} \lambda_k |k\rangle_{2}
\otimes
|k\rangle_{3},
\end{equation}
where the 
nonnegative real Schmidt coefficients are put in decreasing order:
$\lambda_0 \ge \lambda_1 \ge \ldots \lambda_{d-1} \ge 0$.
Using this basis, we may write each of the vectors $|\Phi_r \rangle_{12}$
as
\begin{equation}
|\Phi_r \rangle_{12} = \sum_{k=0}^{d-1}
 |\phi_r^k\rangle_{1} 
\otimes
|k\rangle_{2} ,
\end{equation}
where the vectors $|\phi_r^k\rangle_{1}$ are not
necessarily normalized. Applying this representation, the expression
for the mean fidelity takes the form:
\begin{equation}
\bar{f} = \int \text{d}\psi \sum_{rs}
\left|\sum_{k=0}^{d-1}
\lambda_k \langle \psi | \hat{B}_{rs} |k \rangle \langle
\phi_r^k | \psi
\rangle
\right|^2,
\end{equation}
where all scalar products are now taken in a single-particle Hilbert space.
This allows us to drop the indexes labelling the particles.
The fact that the operators $|\Phi_r\rangle_{12} \langle \Phi_r |$ form
a decomposition of unity implies the following conditions on
$|\phi_r^k\rangle$:
\begin{equation}
\label{Eq:PhiCond}
\sum_{r} |\phi_r^k\rangle \langle \phi_r^l | = \delta_{kl} \hat{\openone}.
\end{equation}
The constraints imposed on the operators $\hat{B}_{rs}$ are given
by $\sum_{s} \hat{B}_{rs}^{\dagger} \hat{B}_{rs} = \hat{\openone}$.
Our task is now to optimize the expression for the mean fidelity $\bar{f}$
over all possible measurements on Alice's side, and transformations performed
by Bob.

We shall start by deriving an upper bound
on the mean fidelity of teleportation
using the state $|\text{tele}\rangle_{23}$. For
this purpose, let us define the vectors $|u_r^k \rangle$ such that:
\begin{equation}
\label{Eq:Easy}
\sum_{k=0}^{d-1} \lambda_k |k\rangle \langle \phi_r^k | 
=
\sum_{k=0}^{d-1} |u_r^k \rangle \langle k | 
.
\end{equation}
The vectors $|u_r^k \rangle$ are uniquely defined by decomposing
$\langle \phi_r^k |$ in the basis $\langle k |$ and collecting
all the terms multiplying each of $\langle k |$'s.
The mean fidelity can be now represented as:
\begin{eqnarray}
\bar{f} & = & 
\sum_{rs} \int \text{d} \psi
\left| 
\sum_{k=0}^{d-1}
\langle \psi | \hat{B}_{rs} | u_r^k \rangle
\langle k | \psi \rangle
\right|^2
\nonumber \\
 & = & \sum_{rs} \sum_{k,l=0}^{d-1}
\langle u_r^k | \hat{B}^\dagger_{rs} \hat{M}_{kl}
\hat{B}_{rs} | u_r^l \rangle
,
\label{Eq:fr}
\end{eqnarray}
where the operators $\hat{M}_{ij}$ are given by the following integrals
over the space of pure states $|\psi\rangle$:
\begin{equation}
\label{Eq:Mkl}
\hat{M}_{kl} = \int \text{d} \psi \,
\langle \psi | k \rangle \langle l | \psi \rangle
\, |\psi\rangle \langle
\psi|
=
\frac{1}{d(d+1)} (\delta_{kl} \hat{\openone} + | k \rangle \langle l |).
\end{equation}
The second explicit form of $\hat{M}_{kl}$ is derived in the Appendix.
Inserting this representation
for the operators $\hat{M}_{kl}$ into
Eq.~(\ref{Eq:fr}), we can reduce the expression for $\bar{f}$ to the form:
\begin{equation}
\label{Eq:fr=16}
\bar{f} = \frac{1}{d(d+1)} \sum_{r}
\left( \sum_{k=0}^{d-1} \langle u_r^k | u_r^k \rangle
+
\sum_{s} \left| 
\sum_{k=0}^{d-1}
\langle k | \hat{B}_{rs} | u_r^k \rangle 
\right|^2 \right) .
\end{equation}
The first sum over $k$ can be transformed with the
help of Eq.~(\ref{Eq:Easy}) multiplied by its hermitian
conjugate:
\begin{eqnarray}
\sum_{k=0}^{d-1} \langle u_r^k | u_r^k \rangle
& = &
\text{Tr} \left( \sum_{k,l=0}^{d-1} |k \rangle \langle u_r^k | u_r^l
\rangle \langle l |  \right)
\nonumber \\
& = &
\text{Tr} \left( 
\sum_{k,l=0}^{d-1} \lambda_k \lambda_l |\phi_r^k \rangle
\langle k | l \rangle \langle \phi_r^l | 
\right)
 = \sum_k \lambda_k^2 \langle \phi_r^k | \phi_r^k \rangle .
\nonumber \\
& &
\end{eqnarray}
Furthermore,
with the help of the same identity given in Eq.~(\ref{Eq:Easy}), we may
convert the expression in the squared modulus in Eq.~(\ref{Eq:fr=16})
to the form:
\begin{eqnarray}
\sum_{k=0}^{d-1} \langle k | \hat{B}_{rs} | u_r^k \rangle 
& = &
\text{Tr} \left(\hat{B}_{rs} \sum_{k=0}^{d-1} |u_r^k \rangle \langle k|
\right)
\nonumber \\
& = & \text{Tr} 
\left(
\hat{B}_{rs} \sum_{k=0}^{d-1}
\lambda_k | k \rangle \langle \phi_r^k | 
\right)
  = 
\sum_{k=0}^{d-1}
\lambda_k \langle \phi_r^k | \hat{B}_{rs} |k \rangle .
\nonumber \\
& &
\end{eqnarray}
Thus we have:
\begin{equation}
\label{Eq:fr=16-2}
\bar{f} = \frac{1}{d(d+1)} \left( \sum_{r}
\sum_{k=0}^{d-1} \lambda_i^2 \langle
\phi_r^k | \phi_r^k \rangle+ \sum_{rs}
\left| \sum_{k=0}^{d-1}
\lambda_k \langle \phi_r^k | \hat{B}_{rs} | k \rangle
\right|^2 \right) .
\end{equation}
The first term in the above expression can be easily calculated using
the condition defined in Eq.~(\ref{Eq:PhiCond}), which implies that:
\begin{equation}
\label{Eq:FirstSum}
\sum_r \langle \phi_r^k | \phi_r^k \rangle
= \text{Tr} \left( \sum_r | \phi_r^k \rangle \langle \phi_r^k |
\right) = \text{Tr} \, \hat{\openone} = d,
\end{equation}
and consequently the double sum over $r$
and $k$ yields $d\sum_{k=0}^{d-1}
\lambda_k^2 =d$.
In order to estimate the second term in Eq.~(\ref{Eq:fr=16-2})
we will use the inequality
\begin{equation}
\label{Eq:Triangle}
\sum_{\alpha=1}^{N}\left|\sum_{k=1}^{M} x_{k\alpha} \right|^2 \le 
\left( \sum_{k=1}^{M} \sqrt{\sum_{\alpha=1}^{N} |x_{k\alpha}|^2} 
\right)^2
\end{equation}
valid for arbitrary  complex numbers $x_{k\alpha}$. This is simply
the triangle inequality for $M$ complex
$N$-dimensional vectors ${\bf x}_k = (x_{k1},\ldots,
x_{kN})$ with the standard quadratic
norm $||{\bf x}_k||^2 = \sum_{\alpha=1}^{N}
|x_{k\alpha}|^2$. When all the vectors
${\bf x}_{k} \neq 0$, the equality sign in Eq.~(\ref{Eq:Triangle})
holds if and only if there exist $M$ strictly positive numbers $a_1,\ldots,
a_{M}$ such that $a_l {\bf x}_k = a_k {\bf x}_l$ for every pair $k,l$.

With the help of the triangle inequality,
we can easily find an upper bound for the sum over
$rs$ in Eq.~(\ref{Eq:fr=16-2}):
\begin{equation}
\label{Eq:Ineq}
\sum_{rs} \left| 
\sum_{k=0}^{d-1}
\lambda_k \langle \phi_r^k | \hat{B}_{rs} |k \rangle
\right|^2 
 \le 
\left( \sum_{k=0}^{d-1} \lambda_k  \sqrt{ \sum_{rs} | 
\langle \phi_r^k | \hat{B}_{rs} |k \rangle |^2}
\right)^2
.
\end{equation}
The sum over $rs$ under square root in the above expression
can be estimated by
\begin{eqnarray}
\sum_{rs} |              
\langle \phi_r^k | \hat{B}_{rs} |i \rangle |^2
& = &
\sum_{r}
\langle \phi_r^k | \phi_r^k \rangle 
\sum_s
\langle k | \hat{B}^{\dagger}_{rs} 
\frac{|\phi_r^k \rangle \langle \phi_r^k |}%
{\langle \phi_r^k | \phi_r^k \rangle}
\hat{B}_{rs} | k \rangle
\nonumber \\
& \le & \sum_{r}
\langle \phi_r^k | \phi_r^k \rangle
\sum_s
\langle k | \hat{B}^{\dagger}_{rs}
\hat{\openone}
\hat{B}_{rs} | k \rangle 
\nonumber \\
& = & \sum_{r} \langle \phi_r^k | \phi_r^k \rangle = d
.
\label{Eq:sumsphiB0}
\end{eqnarray}
In deriving Eq.~(\ref{Eq:sumsphiB0}) we have
implicitly assumed that $|\phi_r^k\rangle \neq
0$, but of course the above inequalities hold also in the case when
$|\phi_r^k\rangle$ is zero.
Thus we finally obtain
the following upper bound on the mean fidelity:
\begin{equation}
\label{Eq:fle}
\bar{f} 
\le  
\frac{1}{d+1} 
\left[ 1 + \left(\sum_{k=0}^{d-1} \lambda_k \right)^2 \right]
.
\end{equation}

We will now analyse necessary and sufficient conditions for a given
teleportation protocol to be an optimal one. This is the case if the
inequality signs in Eqs.~(\ref{Eq:Ineq}) and (\ref{Eq:sumsphiB0})
are replaced by equalities. Let us denote by $m$ the maximum index
for which $\lambda_m$ is nonzero, i.e. $\lambda_{m+1} = \ldots =
\lambda_{d-1} = 0$. Of course, it is sufficient to characterize
the vectors $|\phi_r^k\rangle$ for $k\le m$, and the action of the
operators $\hat{B}_{rs}$ on the subspace spanned by the vectors
$|0\rangle,\ldots,|m\rangle$.

The inequality sign
in Eq.~(\ref{Eq:Ineq}) becomes equality
if and only if there exist $m+1$ nonnegative
numbers $a_0,\ldots,a_{m}$ such that:
\begin{equation}
\label{Eq:FromTriangle}
a_l \lambda_k \langle \phi_r^k | \hat{B}_{rs} | k \rangle = a_{k}
\lambda_l \langle \phi_r^l | \hat{B}_{rs} |l \rangle
\end{equation}
for any pair $k,l \le m$.
Furthermore, equality in
Eq.~(\ref{Eq:sumsphiB0}) takes place if and only if:
\begin{equation}
\label{Eq:Samjuzniewiem}
|\langle \phi_r^k | \hat{B}_{rs} |k \rangle |^2 =
\langle \phi_r^k | \phi_r^k \rangle
\langle k | \hat{B}_{rs}^\dagger \hat{B}_{rs} | k \rangle
.
\end{equation}
Let us note that
for a given $k$ the scalar products
$\langle \phi_r^k | \hat{B}_{rs} | k \rangle$
cannot be identically equal to zero. Otherwise, Eq.~(\ref{Eq:Samjuzniewiem})
implies a contradiction:
\begin{equation}
0 = \sum_{rs} |\langle \phi_r^k | \hat{B}_{rs} | k \rangle|^2
= \sum_{rs} \langle \phi_r^k | \phi_r^k \rangle
\langle k | \hat{B}_{rs}^\dagger \hat{B}_{rs} | k \rangle
= \sum_{r} \langle \phi_r^k | \phi_r^k \rangle = d.
\end{equation}
Consequently, $a_k$ must be strictly positive for $k\le m$,
as discussed after Eq.~(\ref{Eq:Triangle}).
By taking the squared modulus of Eq.~(\ref{Eq:FromTriangle}), making use
of Eq.~(\ref{Eq:Samjuzniewiem}), and performing the summation over $s$ and
$r$ we find $a_l \lambda_k$=$a_k
\lambda_l$ for each pair $k,l$. 
Thus we obtain:
\begin{equation}
\label{Eq:Independent}
\langle \phi_r^k | \hat{B}_{rs} | k \rangle = 
\langle \phi_r^l | \hat{B}_{rs} | l \rangle
\end{equation}
for any $k, l \le m$. Furthermore, 
Eq.~(\ref{Eq:Samjuzniewiem}) implies that
\begin{equation}
\label{Eq:Brs|i>}
\hat{B}_{rs} |k\rangle = \mu_{rsk} |\phi_r^k \rangle
\end{equation}
where $\mu_{rsk}$ are certain complex numbers.
By taking the scalar product of this identity with $\langle \phi_r^k |$,
making use of Eq.~(\ref{Eq:Independent})
we see that the coefficients $\mu_{rsk}$ are independent of $k$:
$\mu_{rsk}=\mu_{rs}$. Then by taking 
the scalar product of Eq.~(\ref{Eq:Brs|i>})
with the hermitian conjugated identity
$\langle l | \hat{B}^\dagger_{rs} = \langle \phi_r^l | \mu_{rs}^{\ast}$,
and performing summation over $s$ we obtain:
\begin{equation}
\langle \phi_r^k | \phi_r^l \rangle \sum_{s} |\mu_{rs}|^2
= 
\sum_{s} \langle k | \hat{B}_{rs}^\dagger \hat{B}_{rs} | l \rangle
= \delta_{kl}.
\end{equation}
As $\sum_{s} |\mu_{rs}|^2 =0 $ would imply that for given $r$ all
operators $\hat{B}_{rs}=0$, the above identity means 
that all the vectors $|\phi_r^k\rangle$ are mutually orthogonal
for $k \le m$. Consequently,  the action
of the operator $\hat{B}_{rs}$ in the subspace
spanned by $|0\rangle, \ldots, |m\rangle$ is equivalent, up to
a multiplicative constant, to the action of
\begin{equation}
\hat{B}_{rs} \propto \sum_{k=0}^{m} |\phi_r^k\rangle \langle k |
\end{equation}
Since for a given $r$
the vectors $|\phi_r^k\rangle$ are mutually orthogonal
and have equal norm, the action of
each of the operators $\hat{B}_{rs}$ is proportional
to the same unitary transformation on the relevant subspace.
Thus, in order to reach the
upper bound for fidelity, it is sufficient for Bob to perform a unitary
transformation described by the following operation
on the subspace spanned by
$|0\rangle, \ldots |m\rangle$:
\begin{equation}
\hat{B}_r = \frac{1}{\sqrt{\langle \phi_r^0 | \phi_r^0
\rangle}} \sum_{k=0}^{m}
|\phi_r^k\rangle \langle k |.
\end{equation}
Necessary and sufficient
conditions for Alice's measurement to be optimal are given by
Eq.~(\ref{Eq:PhiCond}) and the requirement that for $k\le m$ all
vectors $|\phi_r^k \rangle$ have equal norm and are mutually orthogonal.
It is straightforward to check that these conditions are fulfilled
by the standard teleportation protocol \cite{BennBrasPRL93} described
by
\begin{equation}
|\phi^{k}_{r=p+qd} \rangle
=
e^{2\pi i kp/d}|(k + q) \, \text{mod} \, d \rangle
\end{equation}
where $p,q=0,\ldots,d-1$ and the index $r$ runs from $0$ to $d^2-1$.
Consequently, the standard teleportation protocol with appropriately
adjusted bases saturates the upper bound on the mean fidelity derived
in Eq.~(\ref{Eq:fle}).

Of course, use of a nonmaximally entangled state makes the teleportation
imperfect. However,
suppose that Alice would like to use the result of her measurement
to estimate the quantum state which has been teleported. Thus, for 
each outcome $r$ of her measurement, she would like to assign a state
$|\psi_r^{\text{est}}\rangle$, which is her guess for the teleported
state. 
This state can be represented as a result of a unitary
transformation $\hat{U}_r$ performed on a reference state $|0\rangle$:
\begin{equation}
|\psi_r^{\text{est}}\rangle = \hat{U}_r |0\rangle.
\end{equation}
Given the input state $|\psi\rangle$, the probability that Alice's
measurement yields the outcome $r$ equals $\sum_{k=0}^{d-1}
\lambda_k^2 | \langle \phi_r^k | \psi  \rangle|^2$. The fidelity
of the corresponding estimate is then $|\langle \psi |
\psi_r^{\text{est}}\rangle|^2 =
|\langle \psi |
\hat{U}_r |0\rangle|^2$.
Thus, the mean fidelity of Alice's estimate is given by:
\begin{equation}
\bar{f}_{\text{est}} 
 =  \sum_{r} \int \text{d}\psi | \langle \psi | 
\hat{U}_r |0\rangle |^2
\sum_{k=0}^{d-1} \lambda_k^2 | \langle \phi_r^k | \psi \rangle
|^2
\end{equation}
Using the invariance of the measure $\text{d}\psi$ with respect to unitary
transformations, we may change the integration according to
$|\psi \rangle \rightarrow \hat{U}_r | \psi \rangle$. This yields:
\begin{eqnarray}
\bar{f}_{\text{est}}
& = & \sum_{r} \sum_{k=0}^{d-1}
\lambda_k^2 | \langle \psi | 0\rangle |^2
| \langle \phi_r^k | \hat{U}_r | \psi \rangle
|^2
\nonumber \\
& = &
\sum_{r} \sum_{k=0}^{d-1} \lambda_i^2 \langle \phi_r^k | \hat{U}_r
\hat{M}_{00} \hat{U}_r^\dagger | \phi_r^k \rangle
,
\end{eqnarray}
where $\hat{M}_{00}$ is defined in Eq.~(\ref{Eq:Mkl}).
By inserting its explicit form, we obtain that:
\begin{equation}
\bar{f}_{\text{est}}
 = 
\frac{1}{d(d+1)}\left( \sum_{k=0}^{d-1} \lambda_k^2 \sum_{r}
\langle \phi_r^k | \phi_r^k \rangle +
\sum_{k=0}^{d-1} \lambda_k^2 \sum_{r} | \langle \phi_r^k | \hat{U}_r
| 0 \rangle |^2
\right )
\end{equation}
The first double sum over $k$ and $r$ gives $d$, which follows from
Eq.~(\ref{Eq:FirstSum}).  The second sum can be estimated using the fact
that for a given $r$ all the vectors $ | \phi_r^k \rangle $ are orthogonal
and have equal norm for $k\le m$.  Thus, the second sum over $k$ is
maximized if the operator $\hat{U}_r$ maps the vector $|0\rangle$ onto
the subspace spanned by the vectors $| \phi_r^k \rangle$ corresponding
to the maximum $\lambda_k$. As $\lambda_k$ are ordered decreasingly,
we obtain the following upper bound on the estimation fidelity:
\begin{equation}
\label{Eq:fest}
\bar{f}_{\text{est}}
\le \frac{1}{d+1} \left( 1 + \frac{\lambda_0^2}{d} \sum_r \langle
\phi_r^0 | \phi_r^0 \rangle \right)
= \frac{1+\lambda_0^2}{d+1}.
\end{equation}
It is straightforward to see that
the optimal estimation strategy is given by
\begin{equation}
|\psi_r^{\text{est}} \rangle = 
\frac{1}{\sqrt{
\langle \phi_r^0 | \phi_r^0 \rangle}}
| \phi_r^0 \rangle.
\end{equation}
Of course, if several of $\lambda_k$ have the same maximum value,
then Alice can take as a guess any linear combination of the corresponding
vectors $| \psi_r^k \rangle$. 
Let us note that 
the expression
for maximum $\bar{f}_{\text{est}}$ has analogous structure to the
optimal teleportation fidelity $\bar{f}$,
with $\lambda_0$ replacing $\sum_{k=0}^{d-1} \lambda_k$.

It is interesting to
compare two extreme cases: if the state $|\text{tele}\rangle_{23}$
is maximally entangled, the maximum estimation fidelity is $1/d$,
which corresponds to making completely random guesses by Alice. This
is clear, as perfect teleportation with a maximally entangled state cannot
reveal any information on the teleported state.  On the other hand,
if the state $|\text{tele}\rangle_{23}$ is completely disentangled,
the maximum estimation fidelity is $2/(d+1)$, which corresponds to
the optimal state estimation of a $d$-level system from a single
copy \cite{DDimensions}. In this case, the optimal
teleportation strategy reduces to the optimal state estimation procedure,
with Bob generating on his side an imperfect copy according to the
classical message obtained from Alice.

In conclusion, we have derived an upper bound for fidelity of teleportation
using an arbitrary pure bipartite system, and characterized optimal
teleportation protocols. We have also presented an optimal strategy
for estimating the quantum state given result of the measurement
performed in course of teleportation.

The author thanks Prof.\ J.~H.~Eberly for valuable
comments on the manuscript, and
P.~Horodecki and V.~Vedral for helpful discussions. This research
was partially supported by ARO--administered MURI grant No.\
DAAG-19-99-1-0125, and by NSF grant PHY-9415583.

\appendix

\section*{}

In this Appendix, we evaluate integrals in Eq.~(\ref{Eq:Mkl}).
Because of symmetry, it is sufficient to consider
two cases: $i=0,j=0$, and $i=0,j=1$. We will use the following
parameterization of the state vector $|\psi\rangle$ in the
basis $|k\rangle$:
\begin{equation}
|\psi\rangle
= 
\left(
\begin{array}{c}
e^{i\xi} \cos\theta \\
\sin\theta \cos\varphi \\
z_3 \sin\theta \sin\varphi \\
\vdots \\
z_d \sin\theta \sin\varphi
\end{array}
\right),
\end{equation}
where 
$0 \le \xi \le 2\pi$, $0 \le \theta,\varphi \le \pi/2$, and
$z_3,\ldots,z_d$ are complex numbers satisfying $|z_3|^2+\ldots
+ |z_d|^2=1$. 
This parameterization is a straightforward generalization of the
method used in \cite{SchaDAriPRE94}. Following \cite{SchaDAriPRE94},
the invariant volume element in this parameterization
is given by:
\begin{eqnarray}
\text{d}\psi & = &
\frac{(d-1)!}{4\pi^{d-1}}
(\sin\theta)^{2d-3} (\sin\varphi)^{2d-5}
\nonumber \\
& &
\times
\text{d}(\sin\theta) \, \text{d}(\sin\varphi) \,
 \text{d}\xi \,
\text{d}S_{2d-5}
,
\end{eqnarray}
where $\text{d}S_{2d-5}$ is the volume element of the unit sphere
$S_{2d-5}$. 
For the case $i=0,j=0$ all
the off-diagonals elements vanish, and we need to calculate only
two elements: $\langle 0 | \hat{M}_{00} | 0 \rangle =
\int \text{d}\psi \, \cos^4\theta = 2/[d(d+1)]
$, 
and $\langle 1 | \hat{M}_{00} | 1 \rangle =
\int \text{d}\psi \, \sin^2 \theta \, \cos^2 \theta \, \cos^2\varphi
= 1/[d(d+1)]$. 
Due to symmetry, we have $\langle k | \hat{M}_{00} | k \rangle
= 1/[d(d+1)]$ for all $k\neq 0$.
For the operator $\hat{M}_{01}$, the only nonvanishing element is
$
\langle 0 |\hat{M}_{01} | 1 \rangle
=
\int \text{d}\psi \, \sin^2 \theta \, \cos^2 \theta \, \cos^2\varphi
= 1/[d(d+1)]$.

\end{document}